\documentclass[letterpaper, 10 pt, conference]{ieeeconf}
\IEEEoverridecommandlockouts
\usepackage{cite}
\usepackage{flushend}
\usepackage{amsmath,amssymb,amsfonts}
\usepackage{algorithmic}
\usepackage{graphicx}
\usepackage{textcomp}
\usepackage{xcolor}
\usepackage{multirow}
\usepackage{booktabs}
\def\BibTeX{{\rm B\kern-.05em{\sc i\kern-.025em b}\kern-.08em
    T\kern-.1667em\lower.7ex\hbox{E}\kern-.125emX}}
\begin{document}

\title{Immunohistochemistry Biomarkers-Guided Image Search for Histopathology}

\author{Abubakr Shafique$^{1,2}$, Morteza Babaie$^{1,3}$, Ricardo Gonzalez$^{4}$, and H.R. Tizhoosh$^{1,2,3}$
\thanks{$^{1}$Abubakr Shafique, Morteza Babaie, and H.R. Tizhoosh are with Kimia Lab, University of Waterloo, Waterloo, ON, Canada (email:  abubakr.shafique@uwaterloo.ca)}
\thanks{$^{2}$Abubakr Shafique and H.R. Tizhoosh are also affiliated with Rhazes Lab, Department of Artificial Intelligence \& Informatics, Mayo Clinic, Rochester, MN, USA.}
\thanks{$^{3}$Hamid Tizhoosh and Morteza  Babaie are also affiliated with Vector Institute, MaRS Centre, Toronto, Canada.}
\thanks{$^{4}$Ricardo Gonzalez is affiliated with Laboratory Medicine and Pathology, Mayo Clinic, Rochester, MN, USA.}
}
\maketitle

\begin{abstract}
Medical practitioners use a number of diagnostic tests to make a reliable diagnosis. Traditionally, Haematoxylin and Eosin (H\&E) stained glass slides have been used for cancer diagnosis and tumor detection. However, recently a variety of immunohistochemistry (IHC) stained slides can be requested by pathologists to examine and confirm diagnoses for determining the subtype of a tumor when this is difficult using H\&E slides only. Deep learning (DL) has received a lot of interest recently for image search engines to extract  features from tissue regions, which may or may not be the target region for diagnosis. This approach generally fails to capture high-level patterns corresponding to the malignant or abnormal content of histopathology images. In this work, we are proposing a targeted image search approach, inspired by the pathologists' workflow, which may use information from multiple IHC biomarker images when available. These IHC images could be aligned, filtered, and merged together to generate a \emph{composite biomarker image} (CBI) that could eventually be used to generate an attention map  to guide the search engine for localized search. In our experiments, we observed that an IHC-guided image search engine can retrieve relevant data more accurately than a conventional (i.e., H\&E-only) search engine without IHC guidance. Moreover, such engines are also able to accurately conclude the subtypes through majority votes.
\end{abstract}

\section{Introduction}
The emergence of digital pathology (DP) has opened new horizons for histopathology~\cite{kalra2020pan}. Machine Learning (ML) algorithms are able to operate on digitized slides to assist pathologists with different tasks. Whereas ML-involving classification and segmentation methods have obvious benefits for image analysis, image search represents a fundamental shift in computational pathology~\cite{kalra2020yottixel}. Matching the pathology of new patients with already diagnosed and curated cases offers pathologists a new approach to improve diagnostic accuracy through visual inspection of similar cases and a computational majority vote for consensus building~\cite{kalra2020yottixel}.

Pathologists examine tissue slides under a microscope on a regular basis and write diagnostic and prognostic reports based on their visual inspections. The use of immunologic research methodologies in histopathology has resulted in a significant improvement in neoplasm microscopic diagnosis~\cite{kabiraj2015}. Immunohistochemistry (IHC) has become a formidable tool at the pathologist's disposal, despite the fact that histological analysis of hematoxylin and eosin (H\&E) stained tissue sections remain at the core of the discipline~\cite{jordan2002, kabiraj2015, kim2016}. IHC is a technique for detecting specific antigens (proteins) in tissue slices using labeled antibodies that bind with the antigen~\cite{walker2006}. The purpose of staining is to draw attention to the area of interest while also providing contrast against the `background'. Diagnostic IHC tests have been commonly used to investigate and confirm diagnoses by determining a tumor's differentiation route~\cite{kabiraj2015, gown2002}. As oncologists demand more specific diagnoses, IHC has proven to be one of the most essential auxiliary techniques in the characterization of neoplastic disorders in humans~\cite{ramos2008, duraiyan2012}. Simultaneously, archives of digital scans in pathology are progressively becoming a reality as workload increases\cite{kalra2020yottixel}.

\begin{figure*}[t]
\centerline{\includegraphics[width =\textwidth]{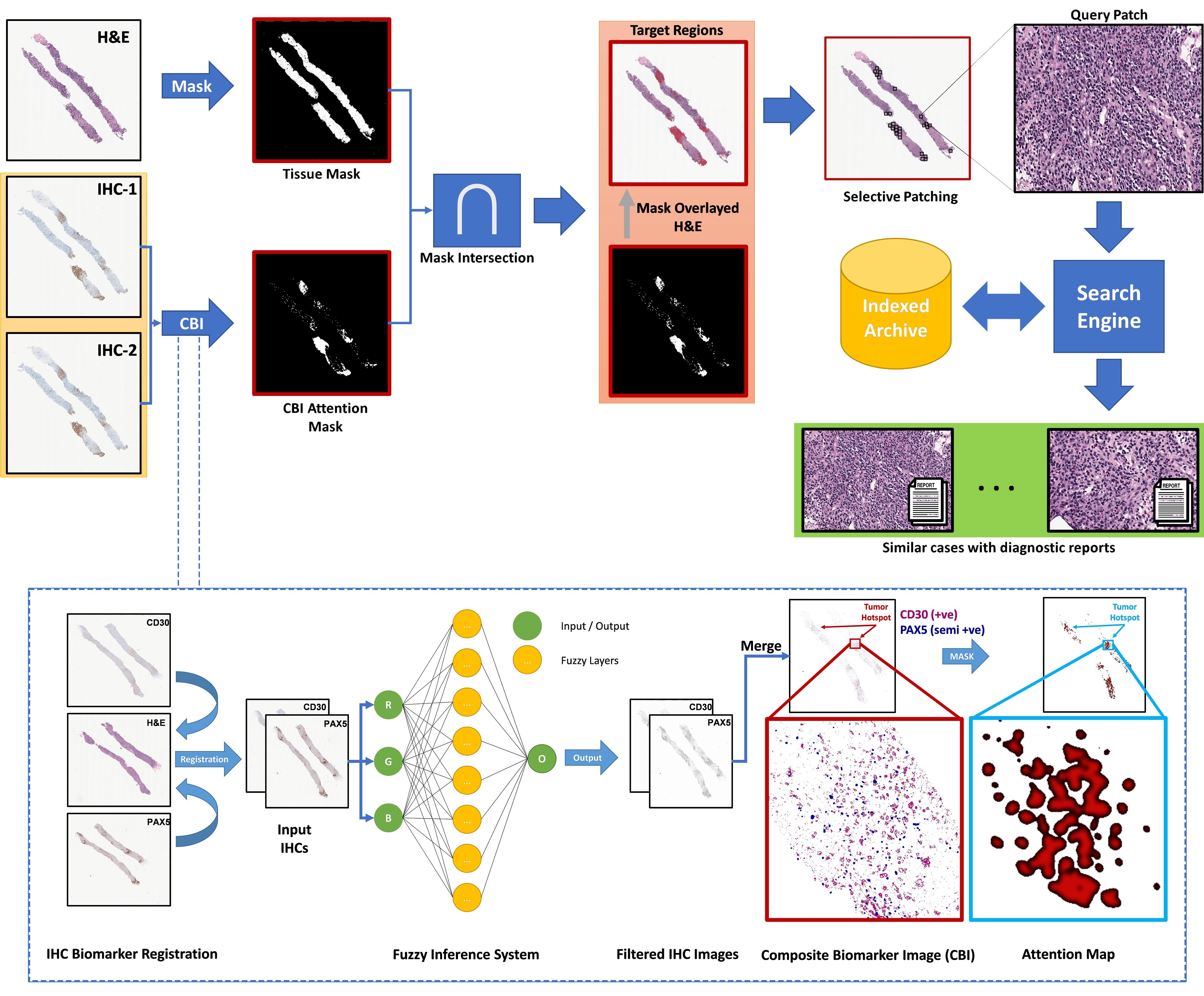}}
\caption{The overall framework for targeted image search. Multiple IHC biomarkers (IHC-1 and IHC-2) are used to generate a CBI-based attention mask, which is used for selective patching. These patches with significant patterns are then indexed by the search engine to accommodate future queries.}
\label{Fig:Methods}
\end{figure*}

Since the last decade, content-based image retrieval (CBIR) has been one of the intensively researched fields in computer vision~\cite{kalra2020pan}. Recently, the use of deep learning (DL) for image search has captured significant attention ~\cite{kieffer2017}. CBIR systems allow the user to search for images that are morphologically and semantically similar to one another. CBIR has numerous real-world applications, and it may be especially valuable for medical images, as linguistic features collected from medical reports may be frequently insufficient to adequately represent the content of the related cases~\cite{doyle2007, dy2003}. The mass medical image archives have traditionally been bundled with textual annotations classified by professionals; however, this technique does not scale well with the ever-increasing demands of DP. While CBIR systems for histopathology have received a lot of attention~\cite{muller2004}, image search and analysis for histopathology has just recently become a focus of research, due to the rise of DP and DL~\cite{kalra2020pan, kalra2020yottixel, hegde2019, shi2017}. Pathologists can access evidence-based knowledge from prior cases through CBIR systems designed specifically for histopathology, allowing them to make decisions more quickly and reliably~\cite{kalra2020yottixel}. Histopathology whole slide images (WSI) are gigapixel files, which generally cannot be processed as a whole and hence are often broken down into numerous ``patches''~\cite{wang2012}. However, the majority of CBIR solutions employ fundamental image features from randomly selected or softly categorized regions, which may or may not be the target region, i.e, the abnormal tissue. This approach generally fails to capture high-level patterns corresponding to the malignant content of histopathology images. The motivation for developing targeted search engines emerged from multiple observations: 1) Not many works have previously incorporated IHC biomarkers in their search engines, 2) Only H\&E-stained WSIs were previously used with random or soft patch selection which may or may not include the embedding from the key suspected regions.

In this work, we propose a targeted image search approach based on the information from the IHC biomarker images. This approach is inspired by pathologists as they use IHC biomarkers for detailed cross-stain analysis, diagnosis, and narrowing down visual inspection to the significant regions with malignant cells. Multiple IHC biomarker images could be aligned, filtered, and merged together to generate a ``composite biomarker image'' (CBI)  to generate an attention map to guide the search engine for better search accuracy.

This paper is organized as follows: Section~\ref{sec:method} describes the algorithms behind the targeted image search and data used in the experiments. The experimental setup and results are described in Section~\ref{Sec:Results}, which are further discussed and concluded in Section~\ref{Sec:Discuss}.

\section{MATERIALS \& METHODS}
\label{sec:method}

Targeted image search, utilizing IHC attention maps, for the histopathology could be a significant upgrade for existing search engines. To make full use of IHC images, we proposed to use CBIs to generate a malignant attention map by combining multiple IHC biomarker images into one (See Fig.~\ref{Fig:Methods}). In ideal scenarios, having multiple medical professionals annotating tumorous regions in many images is quite desirable. But manual delineation is a time-consuming task and subject to user variability.

\begin{figure*}[ht]
\centering{\includegraphics[width = 0.9\textwidth]{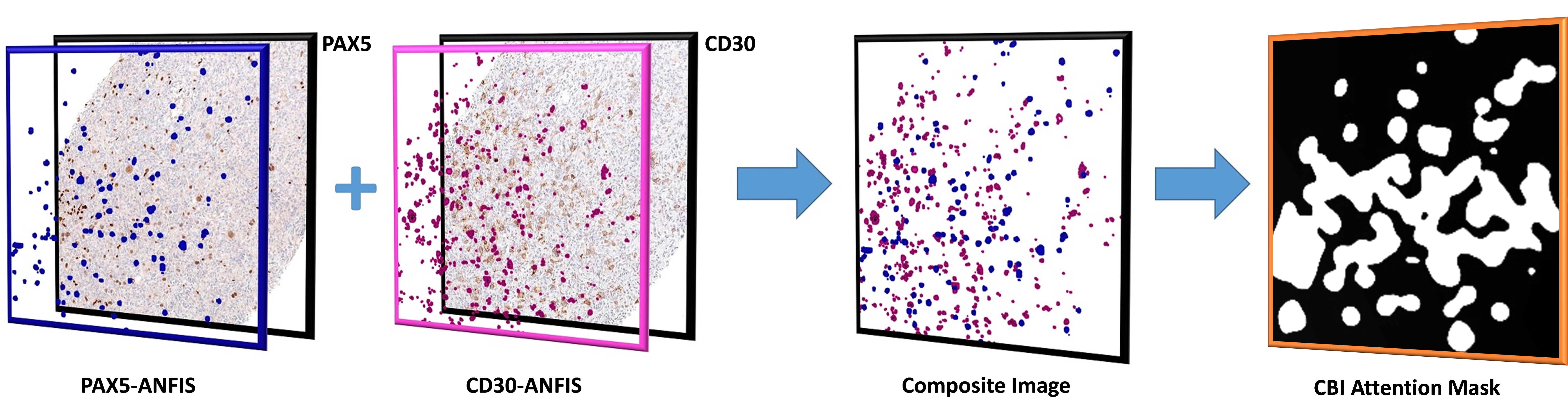}}
\caption{CBI sample produced by the two IHCs (CD30 positive cell population, and PAX5 semi-positive cell population in the same area).}
\label{Fig:CBI}
\end{figure*}

\textbf{IHC Biomarkers \& Alignment -- }
Image alignment is the first step that is crucial for the whole procedure as the morphology and orientation of each cut on different glass slides are often different when looking at IHC and H\&E images. For this reason, we have carefully designed our fully automated non-rigid image registration procedure, which accurately aligns all the biomarkers~\cite{Abubakr2021}. In the registration process, CD30, and PAX5 biomarker WSIs are aligned  to the H\&E-stained WSI. IHC biomarkers CD30 and PAX5 were selected as these are critical for Hodgkin lymphoma (HL)~\cite{ValiBetts2017}. CD30 is a protein that can be detected in T and B lymphocytes (two types of white blood cells). It is a receptor for tumor necrosis factor, a protein that plays a role in cell development and survival~\cite{van2017understanding}. On the other hand, paired box protein, PAX5, is a protein that in humans is encoded by the PAX5 gene. PAX5 is a transcription factor that is required for lymphoid progenitors to commit to the B lymphocyte lineage~\cite{cobaleda2007pax5}. PAX5 has a dual function, inhibiting 'inappropriate' B lineage genes while simultaneously activating B lineage–specific genes~\cite{cobaleda2007pax5}. In the case of classical Hodgkin lymphoma (CHL), it could be identified when there is the presence of CD30 (+) Hodgkin and Reed-Sternberg cells with light expression of PAX5 (B-lymphocytes)~\cite{ValiBetts2017}.

\textbf{Adaptive Neuro-Fuzzy Inference System (ANFIS) -- } After aligning the biomarker images,  ANFIS is used to classify and extract the positive or negative information from the three biomarker images. ANFIS is the combination of neural networks (NN) and FIS which uses NNs learning algorithms to self tune the rules, parameters, and structure of FIS~\cite{Alizadeh2017, Hosseini2012}. For simplicity, we assume a network with three inputs, $r$, $g$, and $b$, and one output, $o$~\cite{jang1992anfis, Hosseini2012}. To design the ANFIS architecture, three \emph{fuzzy if-then rules} based on a first-order Sugeno model are shown below:
\\
\\
\noindent $\textbf{Rule 1:}$ $if$ $r$ $is$ $A_1$ $and$ $g$ $is$ $B_1$ $and$ $b$ $is$ $C_1,$ $then$ $o_1$ $=j_1 r + k_1 g + l_1 b + z_1$\\
\noindent $\textbf{Rule 2:}$ $if$ $r$ $is$ $A_2$ $and$ $g$ $is$ $B_2$ $and$ $b$ $is$ $C_2,$ $then$ $o_2$ $=j_2 r + k_2 g + l_2 b + z_2$\\
\noindent $\textbf{Rule 3:}$ $if$ $r$ $is$ $A_3$ $and$ $g$ $is$ $B_3$ $and$ $b$ $is$ $C_3,$ $then$ $o_3$ $=j_3 r + k_3 g + l_3 b + z_3$
\\
\\
\noindent where $r$, $g$, and $b$ are the inputs, $A_i$, $B_i$, and $C_i$ are the fuzzy sets, $o$ is the output of the fuzzy system, and $j_i$, $k_i$, $l_i$, and $z_i$ are the design parameters that are calculated during the training process.

ANFIS is used to filter the IHC-stained WSIs at pixel level based on thee RGB color, thus taking three inputs as R, G, and B and giving one output $o$ in the range of $0$ to $255$. Here, biomarker specific ANFIS models are trained, for instance with CD30 only positive regions (Class 3, 4, and 5 from table~\ref{tab:Class}), and with PAX5 semi-positive tissues (Class 4 only from table~\ref{tab:Class}), a relevant IHC indication for Hodgkin lymphoma diagnosis. Hence, only IHC-relevant tissue regions will get some grey level and remaining image area will be regarded as irrelevant background. 
For the ANFIS training and testing, 30 RGB values were randomly selected by the pathologist from each class, so in total 180 x 3 (RGB) points were used in our experiment. From 30 points, 25 RGB points were used for the training, and 5 points were used for testing for each class. The range of RGB data for each class for training is shown in Table.~\ref{tab:Class}.

\begin{figure*}[bt]
\centering{\includegraphics[width = 0.9\textwidth]{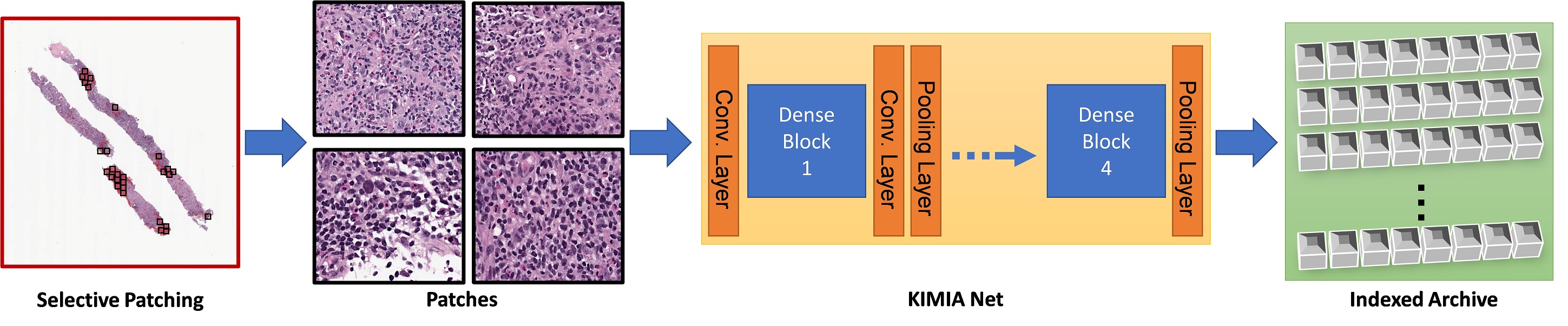}}
\caption{The process of feature extraction and indexing using from the chosen patches using KimiaNet~\cite{riasatian2021}.}
\label{Fig:Target_Search}
\end{figure*}

\begin{table}[htbp]
  \centering
  \caption{\textbf{R, G, and B} input and output data ranges for ANFIS training.}
\begin{tabular}{|c|c|c|l|}
\hline
\multicolumn{3}{|c|}{\begin{tabular}[c]{@{}c@{}}\textbf{Input Range}\end{tabular}} & \multicolumn{1}{c|}{\multirow{2}{*}{\textbf{Class}}} \\ \cline{1-3}
\textbf{R}                        & \textbf{G}                        & \textbf{B}                       & \multicolumn{1}{c|}{}                       \\ \hline
214 - 247                & 214 - 247                & 213 - 247               & 0 = Background                              \\ \hline
24 - 208                 & 44 - 217                 & 79 - 228                & 1 = Blue                                    \\ \hline
63 - 221                 & 65 - 221                 & 77 - 226                & 2 = Gray                                    \\ \hline
163 - 251                & 124 - 218                & 107 - 214               & 3 = Light Brown                             \\ \hline
136 - 192                & 87 - 157                 & 70 - 147                & 4 = Medium Brown                            \\ \hline
37 - 98                  & 0 - 67                   & 0 - 61                  & 5 = Dark Brown                              \\ \hline
\end{tabular}
\label{tab:Class}%
\end{table}

\textbf{Composite Image \& Attention Mask -- }
Each biomarker's grey level image with the relevant tissues are obtained via ANFIS. These images are then processed and combined to create a composite image (see Fig.~\ref{Fig:Methods}). The morphological operations erosion and dilation are used to process all biomarker grayscale images before they are organized and superimposed. In our experiment, we arranged CD30 as the first image, and then PAX5 on top of CD30. A sample CBI image with the two biomarkers is shown in Fig.~\ref{Fig:CBI}. The overlay color of the biomarker images can be chosen by the user, and their overlay order in the CBI image can also be adjusted according to the needs.

Although, annotation of the tumorous regions by the pathologists for training the neural network is required for supervised learning, but it is a cumbersome task subject to observer variability. CBI masks are automated tumor annotation based on the IHC biomarker images (See Fig.~\ref{Fig:CBI}), which are supposed to be similar to a pathologists' annotation when they delineate H\&E images based on their visual clues from IHC slides. The population of targeted cells are focused in IHC biomarker image(s) to get tumorous hotspots (see Fig.~\ref{Fig:Methods} and Fig.~\ref{Fig:CBI}). Therefore, the network learns significant features/patterns more efficiently while training by relaying on targeted tumorous regions.

\textbf{Targeted Image Search -- }
Patches can be matched and retrieved after the target regions have been obtained from the CBI attention mask (See Fig.~\ref{Fig:Methods}). All selected patches from each WSI are now pushed through a pre-trained network namely, KimiaNet (DenseNet-121 architecture), which had been specifically trained and optimized on more than 240,000 histopathological images to distinguish 30 cancer subtypes including normal tissue, benign tumors, and various types of malignant tumors~\cite{riasatian2021}. The last pooling layers or the first connected layers are typically utilized as ``features'' to represent each patch, with the actual output of the network being ignored. In this study, we turn several patch embeddings into a single feature vector of length 1,024 per WSI as the output of the KimiaNet using the median of the minimums approach from Yottixel. (see Fig.~\ref{Fig:Target_Search})~\cite{riasatian2021, kalra2020yottixel}.

\textbf{Data -- }
For the experiments, 15 cases of CHL and 8 cases of nodular lymphocyte predominant Hodgkin lymphoma (NLPHL) were employed, which included H\&E and two IHC biomarkers (CD30 and PAX5). A total of 23 cases, including 69 WSIs and 9,296 patches (size  $300\times300$ at $20\times$ magnification) from H\&E-stain slides were used in this experiment. The WSIs were obtained from the Grand River Hospital ( Kitchener, Ontario, Canada). Specimens were sliced into thin layers and placed onto a paraffin cassette one by one to prepare the glass slides. Furthermore, utilizing a digital scanner, the prepared glass slides were digitized into WSI data at $40\times$ magnification.

\section{Results}
\label{Sec:Results}
\begin{figure}[bt]
\centering{\includegraphics[width = 0.5\textwidth]{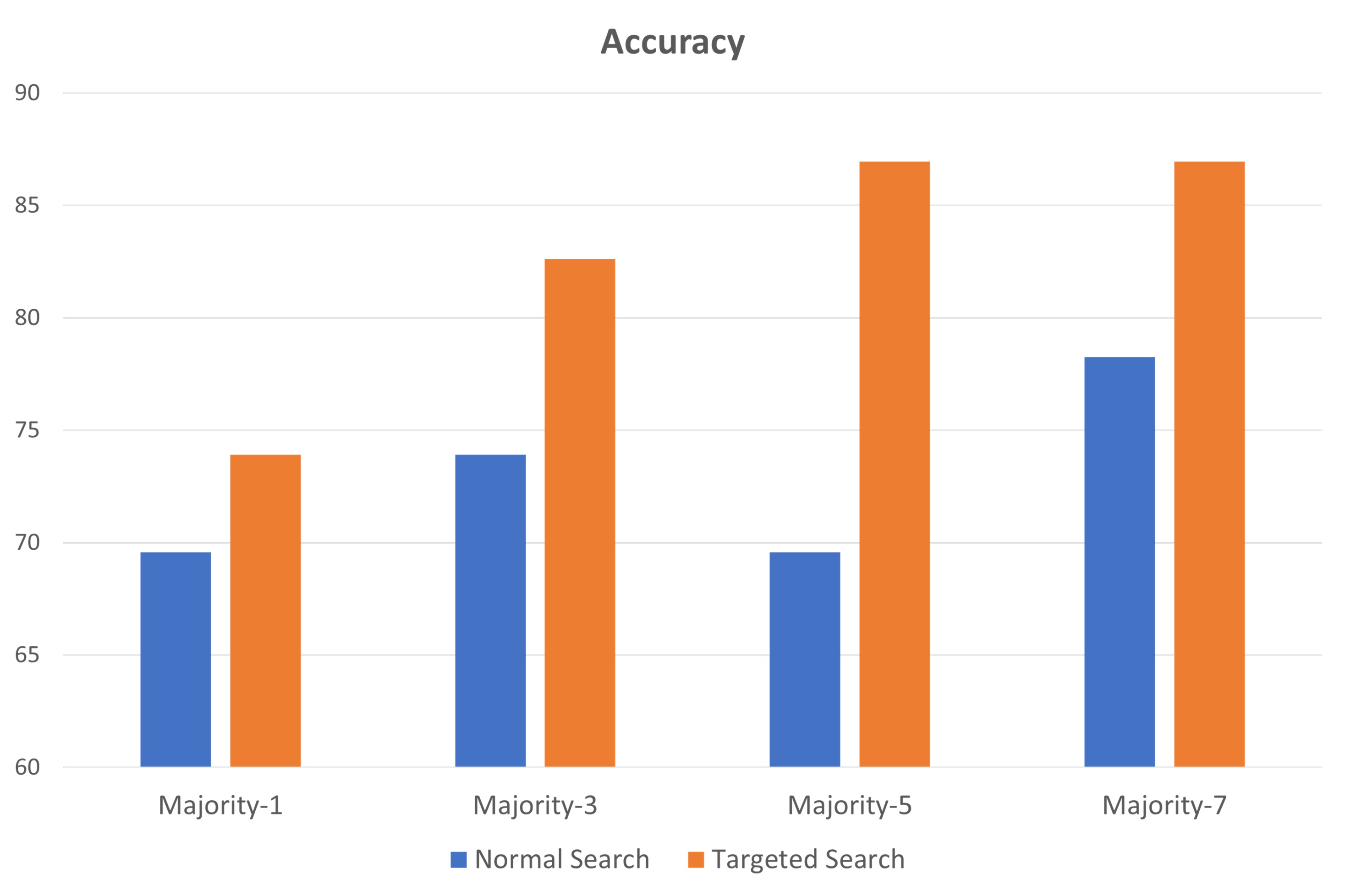}}
\caption{The accuracy of the proposed targeted search (with selective patch selection based on CBI attention mask) and normal search (with random patch selection) according to the majority-1 (top-1), majority-3, majority-5, and majority-7 validation approach.}
\label{Fig:Accuracy}
\end{figure}

In our experiments, the performance of the targeted search (using a CBI attention mask) is compared with the conventional search (without a CBI attention mask) in histopathology. We performed the experiments on a Dell EdgeServer Ra with 2x Intel(R) Xeon(R) Gold 5118 (12 cores, 2.30GHz), 1x Tesla V100 (v-RAM 32 GB), and 394 GB RAM.
The accuracy of image search is evaluated through ``leave-one-patient-out'' validation as the most rigorous technique for small to medium-size datasets~\cite{kalra2020pan}. In this experiment, the majority-$n$ accuracy was determined; only if the majority among $n$ search results were correct, the search considered valid (this is in contrast to the top-$n$ approach in computer vision that regards the attempt successful if \emph{any} of $n$ search results is correct). Precisely, ``correct'' denotes that the majority of detected and retrieved instances correctly identified and matched the tumor type or tumor subtype within a particular diagnostic category. We performed the experiments using H\&E and IHC WSIs from two different types of Hodgkin lymphoma (CHL and NLPHL), retrieving the top one, three, five, and seven WSI patches that match the query. Fig.~\ref{Fig:Accuracy} shows the accuracy and comparison of the proposed targeted search (with CBI attention mask) and normal search (without CBI attention mask) of the top-$n$ results.

For performance evaluation, when searching through the archive to discover the best matches, we employed the $k$-nearest neighbors ($k$-NN) approach to find the top-matched (most similar) embeddings. The best matches, or those closest to the search parameters (based on Euclidean distance), can then be retrieved. for all of the search and matching distance metrics calculations.
 As a result, the feature with the minimum Euclidean distances among the features from two cases (CHL and NLPHL) is the best match. The standard evaluation metrics for verifying the performance of image search and CBIR algorithms are precision, recall, and F1-score when treated as a classifier (see Table~II)~\cite{kalra2020yottixel, riasatian2021}.

\begin{table}
\label{tab:Res}
\caption{Precision, recall, and F1-score for top-1, majority-3, majority-5, and majority-7 search results for the proposed targeted search compared with the normal search.}
\begin{tabular}{cclccc}
\hline
\multicolumn{1}{|c|}{\begin{tabular}[c]{@{}c@{}}Search\\ Type\end{tabular}}                    & \multicolumn{1}{c|}{Tumor}                  & \multicolumn{1}{c|}{\begin{tabular}[c]{@{}c@{}}Top\\ Selection\end{tabular}} & \multicolumn{1}{c|}{Precision}      & \multicolumn{1}{c|}{Recall}         & \multicolumn{1}{c|}{F1-Score}       \\ \hline
\multicolumn{1}{|c|}{\multirow{8}{*}{\begin{tabular}[c]{@{}c@{}}Target\\ Search\end{tabular}}} & \multicolumn{1}{c|}{\multirow{4}{*}{CHL}}   & \multicolumn{1}{l|}{top-1}                                                   & \multicolumn{1}{c|}{\textbf{84.62}} & \multicolumn{1}{c|}{73.33}          & \multicolumn{1}{c|}{\textbf{78.57}} \\
\multicolumn{1}{|c|}{}                                                                         & \multicolumn{1}{c|}{}                       & \multicolumn{1}{l|}{majority-3}                                                   & \multicolumn{1}{c|}{\textbf{86.67}} & \multicolumn{1}{c|}{\textbf{86.67}} & \multicolumn{1}{c|}{\textbf{86.67}} \\
\multicolumn{1}{|c|}{}                                                                         & \multicolumn{1}{c|}{}                       & \multicolumn{1}{l|}{majority-5}                                                   & \multicolumn{1}{c|}{\textbf{100.0}} & \multicolumn{1}{c|}{\textbf{80.00}} & \multicolumn{1}{c|}{\textbf{88.89}} \\
\multicolumn{1}{|c|}{}                                                                         & \multicolumn{1}{c|}{}                       & \multicolumn{1}{l|}{majority-7}                                                   & \multicolumn{1}{c|}{\textbf{100.0}} & \multicolumn{1}{c|}{\textbf{80.00}} & \multicolumn{1}{c|}{\textbf{88.89}} \\ \cline{2-6} 
\multicolumn{1}{|c|}{}                                                                         & \multicolumn{1}{c|}{\multirow{4}{*}{NLPHL}} & \multicolumn{1}{l|}{top-1}                                                   & \multicolumn{1}{c|}{\textbf{60.00}} & \multicolumn{1}{c|}{\textbf{75.00}} & \multicolumn{1}{c|}{\textbf{66.67}} \\
\multicolumn{1}{|c|}{}                                                                         & \multicolumn{1}{c|}{}                       & \multicolumn{1}{l|}{majority-3}                                                   & \multicolumn{1}{c|}{\textbf{75.00}} & \multicolumn{1}{c|}{75.00}          & \multicolumn{1}{c|}{\textbf{75.00}} \\
\multicolumn{1}{|c|}{}                                                                         & \multicolumn{1}{c|}{}                       & \multicolumn{1}{l|}{majority-5}                                                   & \multicolumn{1}{c|}{\textbf{72.73}} & \multicolumn{1}{c|}{\textbf{100.0}} & \multicolumn{1}{c|}{\textbf{84.21}} \\
\multicolumn{1}{|c|}{}                                                                         & \multicolumn{1}{c|}{}                       & \multicolumn{1}{l|}{majority-7}                                                   & \multicolumn{1}{c|}{\textbf{72.73}} & \multicolumn{1}{c|}{\textbf{100.0}} & \multicolumn{1}{c|}{\textbf{84.21}} \\ \hline
\multicolumn{1}{l}{}                                                                           & \multicolumn{1}{l}{}                        &                                                                              & \multicolumn{1}{l}{}                & \multicolumn{1}{l}{}                & \multicolumn{1}{l}{}                \\ \hline
\multicolumn{1}{|c|}{\multirow{8}{*}{\begin{tabular}[c]{@{}c@{}}Normal\\ Search\end{tabular}}} & \multicolumn{1}{c|}{\multirow{4}{*}{CHL}}   & \multicolumn{1}{l|}{top-1}                                                   & \multicolumn{1}{c|}{75.00}          & \multicolumn{1}{c|}{80.00}          & \multicolumn{1}{c|}{77.42}          \\
\multicolumn{1}{|c|}{}                                                                         & \multicolumn{1}{c|}{}                       & \multicolumn{1}{l|}{majority-3}                                                   & \multicolumn{1}{c|}{84.62}          & \multicolumn{1}{c|}{73.33}          & \multicolumn{1}{c|}{78.57}          \\
\multicolumn{1}{|c|}{}                                                                         & \multicolumn{1}{c|}{}                       & \multicolumn{1}{l|}{majority-5}                                                   & \multicolumn{1}{c|}{83.33}          & \multicolumn{1}{c|}{66.67}          & \multicolumn{1}{c|}{74.07}          \\
\multicolumn{1}{|c|}{}                                                                         & \multicolumn{1}{c|}{}                       & \multicolumn{1}{l|}{majority-7}                                                   & \multicolumn{1}{c|}{91.67}          & \multicolumn{1}{c|}{73.33}          & \multicolumn{1}{c|}{81.48}          \\ \cline{2-6} 
\multicolumn{1}{|c|}{}                                                                         & \multicolumn{1}{c|}{\multirow{4}{*}{NLPHL}} & \multicolumn{1}{l|}{top-1}                                                   & \multicolumn{1}{c|}{57.14}          & \multicolumn{1}{c|}{50.00}          & \multicolumn{1}{c|}{53.33}          \\
\multicolumn{1}{|c|}{}                                                                         & \multicolumn{1}{c|}{}                       & \multicolumn{1}{l|}{majority-3}                                                   & \multicolumn{1}{c|}{60.00}          & \multicolumn{1}{c|}{75.00}          & \multicolumn{1}{c|}{66.67}          \\
\multicolumn{1}{|c|}{}                                                                         & \multicolumn{1}{c|}{}                       & \multicolumn{1}{l|}{majority-5}                                                   & \multicolumn{1}{c|}{54.55}          & \multicolumn{1}{c|}{75.00}          & \multicolumn{1}{c|}{63.16}          \\
\multicolumn{1}{|c|}{}                                                                         & \multicolumn{1}{c|}{}                       & \multicolumn{1}{l|}{majority-7}                                                   & \multicolumn{1}{c|}{63.64}          & \multicolumn{1}{c|}{87.50}          & \multicolumn{1}{c|}{73.68}          \\ \hline
\end{tabular}
\end{table}

\section{Conclusions}
\label{Sec:Discuss}
In this paper, we proposed an IHC biomarker-guided image search that identifies the key areas for unlabelled  H\&E WSIs based on the IHC staining patterns. The proposed method can automatically extract the patches from non-annotated H\&E WSIs guided by its corresponding IHC biomarkers WSIs. Patches from these selected regions contain abnormal/malignant tissue, from which the network can extract relevant feature vectors. Consequently, the proposed method improves the accuracy of the search engine compared to the conventional search engines, where they select the patches from WSIs either randomly or based on manual delineations. The results from our experiment confirmed that the proposed IHC-guided search has improved the matching accuracy significantly as we can see in Fig.~\ref{Fig:Accuracy} and Table.~II. 

Future research should focus on the solution for the potentially low availability of IHC biomarker-stained WSIs. Despite the advantages of IHC in identifying the types and subtypes of tumors, IHCs may not be readily available as compared to H\&E slides. Requesting IHC may also result in delays in the diagnostic process.

% References
\bibliographystyle{IEEEtran}
\bibliography{IHC_Guided_Ref.bib}

\begin{thebibliography}{10}
\providecommand{\url}[1]{#1}
\csname url@samestyle\endcsname
\providecommand{\newblock}{\relax}
\providecommand{\bibinfo}[2]{#2}
\providecommand{\BIBentrySTDinterwordspacing}{\spaceskip=0pt\relax}
\providecommand{\BIBentryALTinterwordstretchfactor}{4}
\providecommand{\BIBentryALTinterwordspacing}{\spaceskip=\fontdimen2\font plus
\BIBentryALTinterwordstretchfactor\fontdimen3\font minus
  \fontdimen4\font\relax}
\providecommand{\BIBforeignlanguage}[2]{{%
\expandafter\ifx\csname l@#1\endcsname\relax
\typeout{** WARNING: IEEEtran.bst: No hyphenation pattern has been}%
\typeout{** loaded for the language `#1'. Using the pattern for}%
\typeout{** the default language instead.}%
\else
\language=\csname l@#1\endcsname
\fi
#2}}
\providecommand{\BIBdecl}{\relax}
\BIBdecl

\bibitem{kalra2020pan}
S.~Kalra, H.~R. Tizhoosh, S.~Shah, C.~Choi, S.~Damaskinos, A.~Safarpoor,
  S.~Shafiei, M.~Babaie, P.~Diamandis, C.~J. Campbell \emph{et~al.},
  ``Pan-cancer diagnostic consensus through searching archival histopathology
  images using artificial intelligence,'' \emph{NPJ digital medicine}, vol.~3,
  no.~1, pp. 1--15, 2020.

\bibitem{kalra2020yottixel}
S.~Kalra, H.~R. Tizhoosh, C.~Choi, S.~Shah, P.~Diamandis, C.~J. Campbell, and
  L.~Pantanowitz, ``Yottixel--an image search engine for large archives of
  histopathology whole slide images,'' \emph{Medical Image Analysis}, vol.~65,
  p. 101757, 2020.

\bibitem{kabiraj2015}
A.~Kabiraj, J.~Gupta, T.~Khaitan, and P.~T. Bhattacharya, ``Principle and
  techniques of immunohistochemistry—a review,'' \emph{Int J Biol Med Res},
  vol.~6, no.~3, pp. 5204--5210, 2015.

\bibitem{jordan2002}
R.~C. Jordan, T.~E. Daniels, J.~S. Greenspan, and J.~A. Regezi, ``Advanced
  diagnostic methods in oral and maxillofacial pathology. part ii:
  Immunohistochemical and immunofluorescent methods,'' \emph{Oral Surgery, Oral
  Medicine, Oral Pathology, Oral Radiology, and Endodontology}, vol.~93, no.~1,
  pp. 56--74, 2002.

\bibitem{kim2016}
S.-W. Kim, J.~Roh, and C.-S. Park, ``Immunohistochemistry for pathologists:
  protocols, pitfalls, and tips,'' \emph{Journal of pathology and translational
  medicine}, vol.~50, no.~6, p. 411, 2016.

\bibitem{walker2006}
R.~Walker, ``Quantification of immunohistochemistry—issues concerning
  methods, utility and semiquantitative assessment i,'' \emph{Histopathology},
  vol.~49, no.~4, pp. 406--410, 2006.

\bibitem{gown2002}
A.~Gown, ``Genogenic immunohistochemistry: a new era in diagnostic
  immunohistochemistry,'' \emph{Current Diagnostic Pathology}, vol.~8, no.~3,
  pp. 193--200, 2002.

\bibitem{ramos2008}
J.~A. Ramos-Vara, M.~Kiupel, T.~Baszler, L.~Bliven, B.~Brodersen, B.~Chelack,
  K.~West, S.~Czub, F.~Del~Piero, S.~Dial \emph{et~al.}, ``Suggested guidelines
  for immunohistochemical techniques in veterinary diagnostic laboratories,''
  \emph{Journal of Veterinary Diagnostic Investigation}, vol.~20, no.~4, pp.
  393--413, 2008.

\bibitem{duraiyan2012}
J.~Duraiyan, R.~Govindarajan, K.~Kaliyappan, and M.~Palanisamy, ``Applications
  of immunohistochemistry,'' \emph{Journal of pharmacy \& bioallied sciences},
  vol.~4, no. Suppl 2, p. S307, 2012.

\bibitem{kieffer2017}
B.~Kieffer, M.~Babaie, S.~Kalra, and H.~R. Tizhoosh, ``Convolutional neural
  networks for histopathology image classification: Training vs. using
  pre-trained networks,'' in \emph{2017 Seventh International Conference on
  Image Processing Theory, Tools and Applications (IPTA)}.\hskip 1em plus 0.5em
  minus 0.4em\relax IEEE, 2017, pp. 1--6.

\bibitem{doyle2007}
S.~Doyle, M.~Hwang, S.~Naik, M.~Feldman, J.~Tomaszeweski, and A.~Madabhushi,
  ``Using manifold learning for content-based image retrieval of prostate
  histopathology,'' in \emph{MICCAI 2007 Workshop on Content-based Image
  Retrieval for Biomedical Image Archives: Achievements, Problems, and
  Prospects}.\hskip 1em plus 0.5em minus 0.4em\relax Citeseer, 2007, pp.
  53--62.

\bibitem{dy2003}
J.~G. Dy, C.~E. Brodley, A.~Kak, L.~S. Broderick, and A.~M. Aisen,
  ``Unsupervised feature selection applied to content-based retrieval of lung
  images,'' \emph{IEEE transactions on pattern analysis and machine
  intelligence}, vol.~25, no.~3, pp. 373--378, 2003.

\bibitem{muller2004}
H.~M{\"u}ller, N.~Michoux, D.~Bandon, and A.~Geissbuhler, ``A review of
  content-based image retrieval systems in medical applications—clinical
  benefits and future directions,'' \emph{International journal of medical
  informatics}, vol.~73, no.~1, pp. 1--23, 2004.

\bibitem{hegde2019}
N.~Hegde, J.~D. Hipp, Y.~Liu, M.~Emmert-Buck, E.~Reif, D.~Smilkov, M.~Terry,
  C.~J. Cai, M.~B. Amin, C.~H. Mermel \emph{et~al.}, ``Similar image search for
  histopathology: Smily,'' \emph{NPJ digital medicine}, vol.~2, no.~1, pp.
  1--9, 2019.

\bibitem{shi2017}
X.~Shi, F.~Xing, K.~Xu, Y.~Xie, H.~Su, and L.~Yang, ``Supervised graph hashing
  for histopathology image retrieval and classification,'' \emph{Medical image
  analysis}, vol.~42, pp. 117--128, 2017.

\bibitem{wang2012}
F.~Wang, T.~W. Oh, C.~Vergara-Niedermayr, T.~Kurc, and J.~Saltz, ``Managing and
  querying whole slide images,'' in \emph{Medical Imaging 2012: Advanced
  PACS-Based Imaging Informatics and Therapeutic Applications}, vol.
  8319.\hskip 1em plus 0.5em minus 0.4em\relax International Society for Optics
  and Photonics, 2012, p. 83190J.

\bibitem{Abubakr2021}
A.~Shafique, M.~Babaie, M.~Sajadi, A.~Batten, S.~Skdar, and H.~Tizhoosh,
  ``Automatic multi-stain registration of whole slide images in
  histopathology,'' in \emph{2021 43rd Annual International Conference of the
  IEEE Engineering in Medicine Biology Society (EMBC)}, 2021, pp. 3622--3625.

\bibitem{ValiBetts2017}
E.~Vali~Betts, D.~M. Dwyre, H.-Y. Wang, and H.~H. Rashidi, ``Pax5-negative
  classical hodgkin lymphoma: A case report of a rare entity and review of the
  literature,'' \emph{Case Reports in Hematology}, vol. 2017, p. 7531729, Oct
  2017.

\bibitem{van2017understanding}
C.~Van~der Weyden, S.~Pileri, A.~Feldman, J.~Whisstock, and H.~Prince,
  ``Understanding cd30 biology and therapeutic targeting: a historical
  perspective providing insight into future directions,'' \emph{Blood cancer
  journal}, vol.~7, no.~9, pp. e603--e603, 2017.

\bibitem{cobaleda2007pax5}
C.~Cobaleda, A.~Schebesta, A.~Delogu, and M.~Busslinger, ``Pax5: the guardian
  of b cell identity and function,'' \emph{Nature immunology}, vol.~8, no.~5,
  pp. 463--470, 2007.

\bibitem{Alizadeh2017}
M.~Alizadeh, O.~H. Maghsoudi, K.~Sharzehi, H.~Reza~Hemati, A.~Kamali~Asl, and
  A.~Talebpour, ``Detection of small bowel tumor in wireless capsule endoscopy
  images using an adaptive neuro-fuzzy inference system,'' \emph{Journal of
  biomedical research}, vol.~31, no.~5, pp. 419--427, Sep 2017.

\bibitem{Hosseini2012}
M.~S. Hosseini and M.~Zekri, ``Review of medical image classification using the
  adaptive neuro-fuzzy inference system,'' \emph{Journal of medical signals and
  sensors}, vol.~2, no.~1, pp. 49--60, Jan 2012.

\bibitem{jang1992anfis}
J.~Jang, ``Anfis: Adaptive-network-based fuzzy,'' \emph{IEEE Trans Neural
  Netw}, vol.~3, no. 889, p.~98, 1992.

\bibitem{riasatian2021}
A.~Riasatian, M.~Babaie, D.~Maleki, S.~Kalra, M.~Valipour, S.~Hemati,
  M.~Zaveri, A.~Safarpoor, S.~Shafiei, M.~Afshari \emph{et~al.}, ``Fine-tuning
  and training of densenet for histopathology image representation using tcga
  diagnostic slides,'' \emph{Medical Image Analysis}, vol.~70, p. 102032, 2021.

\end{thebibliography}

\end{document}